\documentclass[pre,aps,twocolumn,showpacs,floatfix]{revtex4-1}

\usepackage{graphicx}
\usepackage{epsfig}
\usepackage{amssymb,amsmath}
\usepackage{amsthm}
\usepackage{bm}

\usepackage[english]{babel}
\usepackage[utf8]{inputenc}
\usepackage[usenames]{color}
\usepackage{hyperref}
\usepackage[all]{hypcap}
\usepackage[normalem]{ulem}

\DeclareMathOperator{\sgn}{sgn}

\newcommand{\sigmaY}{{\sigma_{\tt Y}}}
\newcommand{\tauOff}{{\tau_{\tt res}}}

\begin{document}

\title{Relaxation in yield stress systems through elastically interacting activated events}

\author{Ezequiel E.~Ferrero}
\affiliation{Universit\'e Grenoble Alpes, LIPHY, F-38000 Grenoble, France}
\affiliation{CNRS, LIPHY, F-38000 Grenoble, France}

\author{Kirsten Martens}
\affiliation{Universit\'e Grenoble Alpes, LIPHY, F-38000 Grenoble, France}
\affiliation{CNRS, LIPHY, F-38000 Grenoble, France}

\author{Jean-Louis Barrat}
\affiliation{Universit\'e Grenoble Alpes, LIPHY, F-38000 Grenoble, France}
\affiliation{CNRS, LIPHY, F-38000 Grenoble, France}

\begin{abstract}
We study consequences of long-range elasticity in thermally assisted dynamics of yield stress materials.
Within a two-dimensinal mesoscopic model we calculate the mean-square displacement and the dynamical structure factor
for tracer particle trajectories.
The ballistic regime at short time scales is associated with a compressed exponential decay in the dynamical
structure factor, followed by a subdiffusive crossover prior to the onset of diffusion.
We relate this crossover to spatiotemporal correlations and thus go beyond established mean field predictions.
\end{abstract}


\maketitle


Relaxation of the microscopic structure in glasses, and more generally in soft yield stress materials,
is a topic of long-standing interest and great complexity.
Broad ranges of time, energy and length scales are involved, together with nonequilibrium aspects such
as aging and a strong dependence on the sample preparation scheme.
As a result, no unique scenario has emerged to describe the relaxation of density fluctuations in systems
that, quenched from a liquid into a glassy (solid) state, still display internal dynamics strong
enough to produce structural relaxation on a measurable time scale.
The complexity of the relaxation is usually quantified by the manner in which it deviates from
exponential.
In many cases, stretching, corresponding to a broad distribution of relaxation times, is observed.
However, the opposite situation of compressed relaxation (i.e., faster than exponential) has
emerged in the last years as a new paradigm.
In this work, we confirm through the numerical study of a simplified model that this  behavior can
result from thermally activated plastic events akin to the shear transformations observed in yield stress
solids undergoing external deformation.

A milestone in the experimental analysis of the relaxations processes at hand has been achieved
by a series of dynamic light-scattering experiments on colloidal
gels~\cite{CipellettiMaBaWe-PRL2000, RamosCi2001, CipellettiRaMaPiWePaJo-FD2003,
CipellettiRa-JPCM2005, DuriCi-EPL2006}.
More recently, x-ray photon correlation spectroscopy~\cite{MadsenLeGuSpCz-NJP2010}
has been used to study slow dynamics, not only in supercooled liquids~\cite{CaronnaChMaCu-PRL2008},
colloidal suspensions~\cite{AngeliniZuFlMaRuRu-SM2013} and
gels~\cite{OrsiCrBaMa-PRL2012, OrsiRuChPuRuBaRiCr-PRE2014},
but also in hard amorphous materials like metallic glasses~\cite{RutaChMoCiPiBrGiGo-PRL2012, RutaBaMoCh-JCP2014}.
A common denominator of these experiments is the decay of the dynamical structure factor 
as a compressed exponential in time $t$ and scattering vector $q$:
$f(q,t) \sim \exp[-(t/\tau_f)^\gamma]$, with $\tau_f \sim q^{-n}$, $n\simeq 1$ and shape parameter $\gamma>1$.
The observed dynamics was \textit{a priori} unexpected, not only because of the faster than exponential decay 
of the correlations but also because of the ballistic dynamics contrasting the usual diffusive behavior
($\tau_f \sim q^{-2}$) in molecular dynamics simulations of glassy systems (see, for example,~\cite{ElMasriBeCi-PRE2010}).
Simulations on a gel-former model~\cite{SawElKoSa-PRL2009, SawElKoSa-JCP2011} have shown a
compressed exponential decay of $f(q,t)$, but this was explained as a trivial effect of Newtonian dynamics.

Originally, a heuristic explanation for the observed phenomena was based on the {\it syneresis} of a gel:
the gel shrinks locally and the inhomogeneity acts as a dipole force with a long range-elastic effect~\cite{CipellettiMaBaWe-PRL2000}.
A simple mean-field model approach~\cite{BouchaudPi-EPJE2001, BouchaudPi-EPJE2002, Bouchaud-InBook2008} further
encouraged the view that anomalous relaxation has its origin in elasticity effects,
and stressed its dependence on the time scales considered.
On the other hand, this approach was reported to fail in emulating a $q$ dependence of $\gamma$
observed in experiments~\cite{DuriCi-EPL2006, CaronnaChMaCu-PRL2008, OrsiRuChPuRuBaRiCr-PRE2014}.
Independently, a phenomenological continuous time random walk (CTRW) model with L\'evy flights was introduced~\cite{DuriCi-EPL2006}.
It was used to fit the crossover, with $q$, between compressed and noncompressed behaviors.
However, the assumed Poissonian distribution for the number of events
and the particular power-law distribution for the displacements have never been confirmed.

In this Letter we propose a novel minimalistic model
at the mesoscale for thermally activated relaxation dynamics in yield stress materials. 
After introducing in the first part the main assumptions and the model description,
we validate our results against mean-field predictions for elasticity effects
in the relaxation.
Later, we go beyond mean field and reveal effects due to correlations in the
dynamics that give rise to new interesting phenomena.
We conclude with a discussion of our results and their impact on the
understanding of recent experimental findings.


{\it The model --}
Our model for the coarse-grained relaxation dynamics
is based on two main ingredients: thermally activated yield events (plastic rearrangements)
and a long-range elastic response of the surrounding medium. 
To simplify further an \textit{a priori} tensorial description, we assume rearrangements occurring
along only one axis, such that we can describe the system with scalar quantities for local stresses and
deformations~\cite{PicardAjLeBo-EPJE2004, PicardAjLeBo-PRE2005}.

The yielding of a site leads to a rearrangement with a local deformation rate
given by $\partial_t {\epsilon}^{pl}({\bm r},t) = n({\bm r},t) \varepsilon({\bm r},t)/(2\tau) $,
where $\tau=1$ is a mechanical relaxation time defining our time scale, and $n({\bm r},t)$ is a 
local ``state variable'' indicating whether a site is yielding ($n=1$) or not ($n=0$).
The typical strain $\varepsilon$ caused by a rearrangement is given by $\varepsilon({\bm r},t)=\pm\varepsilon_0$
integrated over the average duration of an event, depending 
only on a sign according to the yielding direction~\cite{BouttesVa-ACP2013}.

The response of the surrounding medium is modeled by using the Eshelby theory of elasticity~\cite{Eshelby1957}.
If $G({\bm r},{\bm r'})$ is the solution for the far field elastic response to a deformed inclusion,
an overdamped dynamics for the coarse-grained scalar stress field $\sigma({\bm r},t)$ reads
\begin{equation}\label{eq:eq_of_motion}
\partial_t \sigma({\bm r},t) = 2\mu \int d{\bm r'} G({\bm r},{\bm r'})\partial_t{\epsilon}^{pl}({\bm r'},t) ,
\end{equation}
where $\mu$ is the elastic modulus.
The propagator for an infinite system in polar coordinates~\cite{Eshelby1957, PicardAjLeBo-EPJE2004} is
$G^\infty(r,\theta) = 2\cos(4\theta)/\pi r^2$.
We discretize Eq.~(\ref{eq:eq_of_motion}) in time and space, on a square lattice
(typical linear sizes $L=2^8,2^{9}$) with periodic boundary conditions.
We solve the evolution of $\sigma({\bm r},t)$ using a pseudospectral method and the discretized propagator in 
Fourier space~\cite{PicardAjLeBo-EPJE2004, PicardAjLeBo-PRE2005}.

The stochastic activation rules for $n({\bm r},t)$ depend only on the local stress $\sigma$, 
two symmetric yield stresses $\sigmaY=\pm \sigma_0$, that mimic the local energy barrier to
yield, and the temperature $T$.
Sites with $|\sigma|>|\sigmaY|$ become immediately active ($n:0\to1$), while the activation probabilities for
sites with $|\sigma|<|\sigmaY|$ read
\begin{equation}\label{eq:activationrule}
 p_{\tt on (\pm)} =  \Gamma_0 \exp\left[\frac{-(\sigmaY^2\mp\sgn(\sigma)\sigma^2)}{2\kappa T}\right].
\end{equation}
Independent of the stress value, active sites deactivate ($n:1\to0$) at a fix rate, 
$p_{\tt off} = \tauOff^{-1}$.
$\Gamma_0$ is an attempt frequency and $\tauOff$ the typical duration of a restructuring event. 
All characteristic times are chosen to be equal to $\tau=1/\Gamma_0=\tauOff=1$, unless otherwise specified.
$\kappa=\mu V_0^{-1}k_B$ is a unitary constant that provides the right magnitudes and $\sigma_0=1$ defines the stress units.
The lack of disorder in the local yield stress leads to steady state dynamics without aging effects.
This simplifies enormously the analysis of the dynamics. 
For each temperature $T$, we reach a unique steady-state
characterized by stress fluctuations around zero with an approximately Gaussian distribution.

To establish an analogy with experiments we need to introduce particles in the model.  
We calculate at each time step the vectorial displacement field ${\bm u}({\bm r},t)$ associated with
the discretized plastic strain field~\cite{PicardAjLeBo-EPJE2004} and introduce tracer particles that 
follow this field, with no further interactions, mimicking the underlying particle dynamics~\cite{MartensBoBa-PRL2011}.
For example, considering a single event at the origin the resulting displacement field reads 
$\mathbf{u}(\mathbf{r})=({2 \ell^2} \varepsilon_0 xy/\pi r^4)\mathbf{r}$, with $\ell$ the lattice parameter and unit of length.
The typical strain change due to a rearrangement is expected to be material dependent.
We choose here $\varepsilon_0=1$, a rather large value that enhance spatial correlations and
make the resulting effects more visible.
Qualitatively, the observed phenomena persist for smaller $\varepsilon_0$, although possibly less noticeable.

The observed dynamical features can essentially be separated in two categories:
On one hand, part of the phenomenology is purely due to the characteristic spatial decay of the elastic
propagator and can easily be captured in mean-field descriptions.
On the other hand, the presence of spatiotemporal correlations in the system leads to the prediction of 
a new dynamical regime.
Yet, before going into the discussion of the correlation effects
we show the compatibility of our model with the mean-field predictions~\cite{BouchaudPi-EPJE2001, BouchaudPi-EPJE2002, Bouchaud-InBook2008}.

{\it Elasticity effects --}
In order to compare our {\it thermal model} with the mean-field results,
we implement independently an analogous system with the sole difference
of having randomly activated sites; a Poissonian
rule for activation instead of the stress-dependent rule (\ref{eq:activationrule}).
In this second {\it random model}, we control the activity, that is,
the number of events per unit time.
We measure the mean activity $a = \left<\frac{1}{N} \sum_i n_i\right>$
(time average) in the thermal model and plug it in the random model as
a parameter to compare equivalent systems\footnote{For low enough temperatures
the mean activity decreases exponentially in $E/T$ with $E \sim 0.48$.}.

\begin{figure}[!thp]
\begin{center}
\includegraphics[width=\columnwidth,clip]{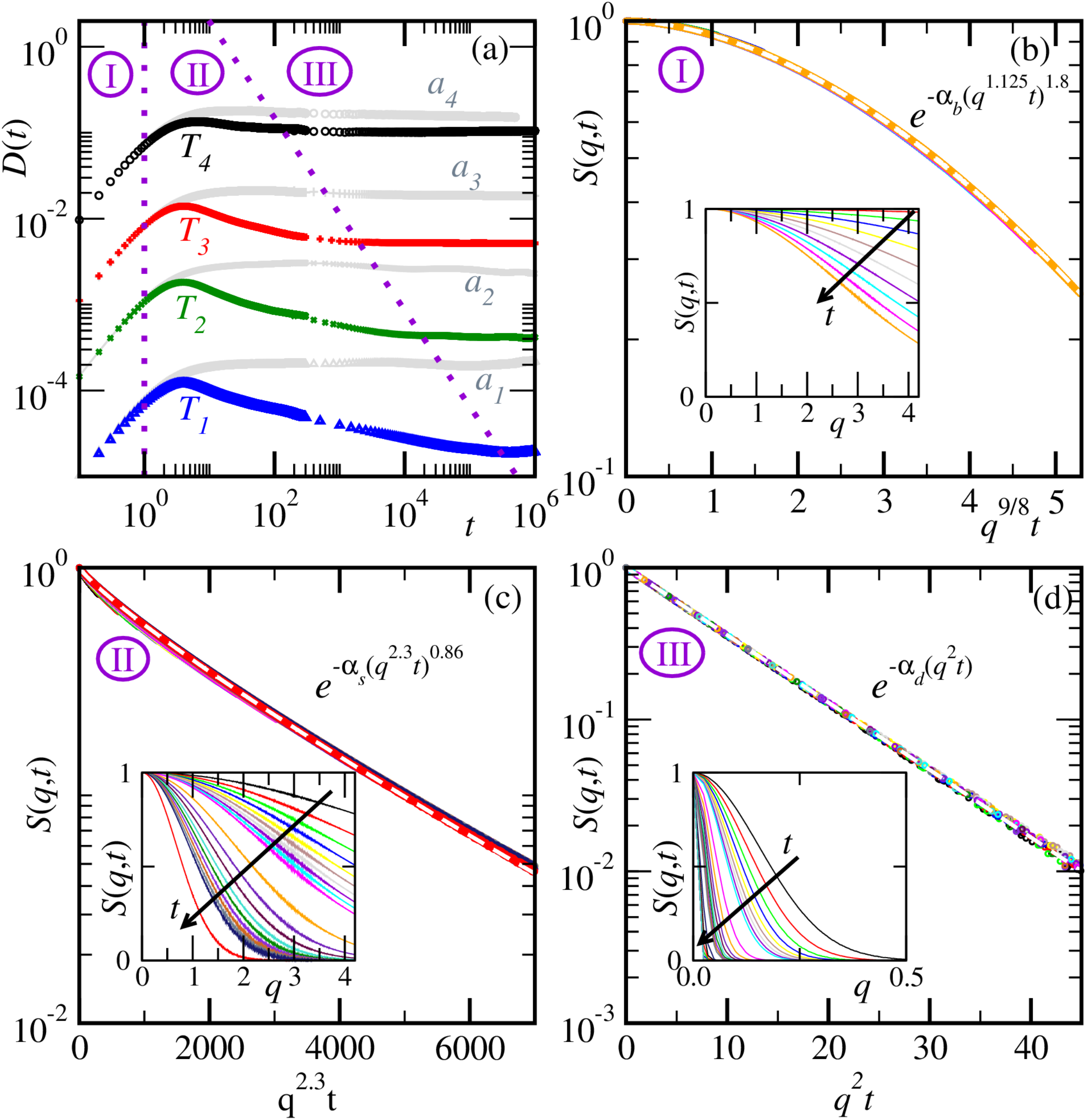} 
\end{center}
\caption{\label{fig:diffusion-and-structure-factor}
{\it Tracer particle dynamics} --
Numerical measurements of self-diffusion coefficients and dynamical structure factors in the steady state.
(a) Diffusion coefficient $D=\langle (\Delta r)^2 \rangle/(4t)$ as a function of time $t$ for different temperatures
$T_i=0.05, 0.07, 0.1, 0.2$.
Pointed lines guide the eye to distinguish three different dynamical regions,
(I) ballistic, (II) crossover, (III) diffusive.
The gray curves show results for the random model (see text) with a mean activity
$a_i \approx 3.1 e^{-0.48 / T_i} $ corresponding to the same temperatures.
(b) Dynamical structure factor $S(q,t)$ for $T=0.2$ as a function of $q^{1.125}t$ for time intervals
corresponding to the ballistic regime (I), fitted by an compressed exponential with shape parameter
$\gamma\approx1.8$ and $\alpha_b\approx 0.07$ (dashed line).
The inset shows the raw data.
(c) Dynamical structure factor $S(q,t)$ for $T=0.07$ as a function of $q^{2.3}t$ for time intervals
corresponding to the crossover regime (II), fitted by a stretched exponential with shape parameter
$\gamma\approx0.86$ and $\alpha_s\approx 0.0015$ (dashed line).
The inset shows the raw data.
(d) Dynamical structure factor $S(q,t)$ for $T=0.2$ as a function of $q^2t$ for time intervals
corresponding to the diffusive regime (III), fitted by a pure exponential with shape parameter
$\gamma = 1$ and $\alpha_D\approx 0.1$ (dashed line).
The inset shows the raw data.
}
\end{figure}

In Fig.~\ref{fig:diffusion-and-structure-factor}(a) we compare for both models
the evolution of the diffusion coefficient $D(t)=\langle\Delta r^2 \rangle/(4t)$,
where the mean square displacement $\langle\Delta r^2\rangle$
on a time window $t$ is averaged both over number of tracers (typically $2^{13}, 2^{14}$)
and sliding time $t_0$, for distances $\Delta r_i = |\mathbf{r}_i(t_0+t) - \mathbf{r}_i(t_0)|$
traveled by each tracer $i$.
We observe that the initial ballistic regime (regime I), whose duration is related
with the persistence $\tau_\mathrm{res}$ of the events, does not depend on the model. 
Also, for both dynamics we find a long time diffusive behavior,
only with different values of the diffusion coefficient (regime III).
Although there exists a third intermediate subdiffusive regime in the spatial
model (regime II), we will focus first on the two regimes that can be predicted
by mean-field considerations.

In analogy with experimental measurements,  we calculate the dynamical structure factor,
\begin{equation}
S(q,t)=\frac{1}{M} \left< \left[ \sum_{n=1}^{M} \cos[\mathbf{q}\cdot(\mathbf{r}_n(t+t_0)-\mathbf{r}_n(t_0))] \right]
\right>_{t_0, |\mathbf{q}|=q} \nonumber
\end{equation}
\noindent where $M$ is the total number of tracers, and the brackets average 
over the sliding time window $t_0$ and the different discretized wave vectors that share the same modulus~\footnote{$(q_x,q_y)$
with $q_a=2\pi k/L$, $a\in\{x,y\}$, $k=0,1,\ldots,L-1$.}.
From mean-field considerations~\cite{BouchaudPi-EPJE2001} we expect for the ballistic regime a decay of $S(q,t)$ 
as a compressed exponential with a dimensionality-dependent shape parameter
$\gamma_{2d}=2$ ($\gamma_{3d}=3/2$)~\footnote{In Bouchaud-Pitard's notation we are always in the case where
$D\theta \gg qv_0$, since we disregard friction in the evolution of $\mathbf{u}(\mathbf{r},t)$ and then $D=K/\gamma \to \infty$.
The intermediate compressed regime with $\gamma'_{3d}=5/4$ is squeezed out.}.
If we search for the best fit of the data for $\tau_f\propto q^{-1}$ we find indeed $\gamma\approx 2$,
but the best collapse of the data is achieved for $\tau_f\propto q^{-1.125}$ yielding a fit with $\gamma=1.8$
[see Fig.~\ref{fig:diffusion-and-structure-factor}(b)].
In the diffusive regime, we can collapse the data by plotting $S(q,t)$ as a function of $q^2t$ and
we obtain a pure exponential decay with $\gamma =1$ as expected 
[see Fig.~\ref{fig:diffusion-and-structure-factor}(d)].
Even when we show these results for a particular temperature, they hold for all
the range of analyzed temperatures (and further, also in the equivalent random model);
only the prefactors $\alpha_b$, $\alpha_d$ are $T$ dependent.

\begin{figure}[!thp]
\begin{center}
\includegraphics[width=\columnwidth, clip]{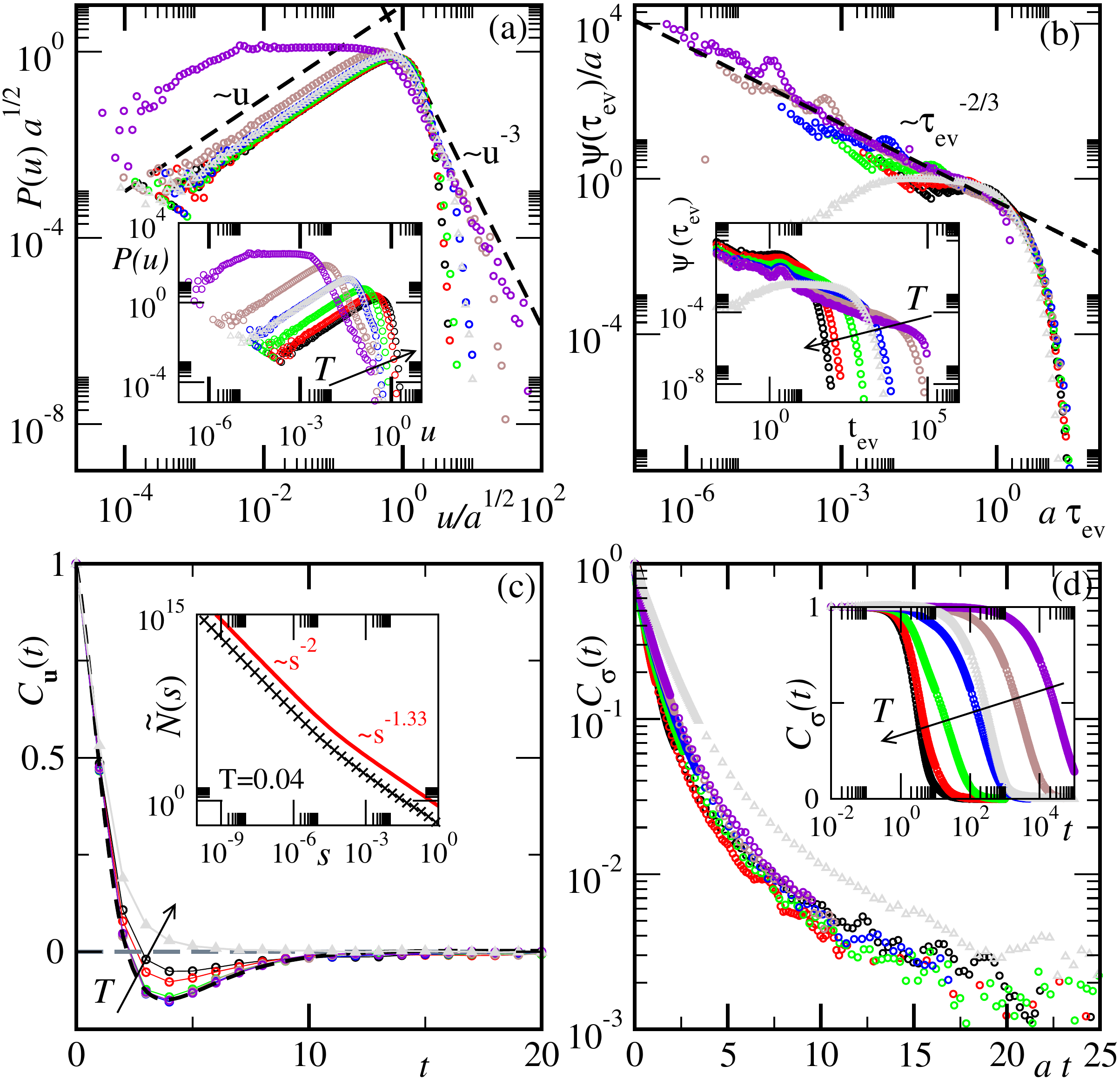} 
\end{center}
\caption{\label{fig:statistical-properties}
{\it Statistical features of the thermally activated dynamics} -- 
In all panels, circles correspond to the thermal model for
temperatures $T_i=0.04, 0.05, 0.07, 0.1, 0.15, 0.2$,
while gray triangles stand for the random model at $a \simeq 0.0034$.
(a)~Distribution of absolute displacements per unit time $u=|{\bm u}|$ of the tracer particles
for different temperatures, rescaled by the square root of the average activity $a(T_i)$.
Dashed lines display power laws.
The inset shows the data without rescaling.
(b)~Rescaled local probability distribution $\psi(\tau_\mathrm{ev})/a(T_i)$ for rescaled
waiting times $a(T_i)\tau_\mathrm{ev}$ between events.
The power-law dashed line serves as a guide to the eye.
The inset shows raw data.
(c)~Two time autocorrelation function $C_{\bm u}(t)$ of the vectorial displacements of
tracer particles.
The inset shows the result of a CTRW model with the Laplace transform of
$\psi(a\tau_\mathrm{ev})$ as an input.
(d)~Two time autocorrelation function $C_\sigma(t)$ of the local stress as a function of the
rescaled time $a(T_i)t$.
The inset shows raw data.
}
\end{figure}

Assuming a long-range elastic response to the local relaxation processes,
we expect the displacement field to decay as $u \sim 1/r^{d-1}$, where $r$ is the distance to the 
event and $d$ the dimensionality of the system.
From a mean-field analysis the distribution of particle displacements
is expected to decay as $P(u)\propto u^{-(2d-1)/(d-1)}$ for large $u$,
yielding for our two-dimensional study $P(u)\propto u^{-3}$, with a finite mean value.
This results directly from the strong elastic response at small distances.
The probability for small displacements on the other hand should grow as $u^{d-1}$, due to the far field 
effect of the propagator.
The crossover between these two regimes should depend on the density of  events, that is, on the activity $a$.
We confirm these scalings within our simulations for low temperatures [see Fig.~\ref{fig:statistical-properties}(a)].
For high temperatures we expect the assumptions of the mean-field description to break down, due to the high density
of events that leads to a screening of the large displacements.

{\it Correlations effects --}
One of the main differences of the thermal model compared  to
random dynamics is the appearance of subdiffusion.
While the random model changes from a ballistic to a pure diffusive behavior
for all activation probabilities, a comparable (same activity) thermal model,
where spatial correlations are allowed to arise,
develops an intermediate subdiffusive regime for low enough temperatures
[regime II in Fig.~\ref{fig:diffusion-and-structure-factor}(a)].

To determine the origin of this effect, we first check
if the tracer displacements are essentially changed
when considering systems with and without spatial correlations.
We find that the distribution for the absolute displacements is not altered
[see Fig.\ref{fig:statistical-properties}(a)].
The change of the dynamics is rather due to negative correlations
in the two-time autocorrelation function of the vectorial displacements
[Fig.\ref{fig:statistical-properties}(c)].
Note that the resolution of the correlation measurement is not sufficient
to determine the extension of the subdiffusive regime, which instead
is seen in the intermittent dynamics of the local rearrangements [Fig.~\ref{fig:statistical-properties}(b)].

Defining $\tau_{ev}$ as the elapsed time between two consecutive activations of the same site,
the resulting distribution $\Psi(\tau_{ev})$ shows a trivial exponential shape
in the random case (expected by construction) but a power-law form with an exponential cutoff
in the thermal model, well scalable with the mean activity in a master curve
$\Psi(a\tau_{ev})\sim (a\tau_{ev})^{-2/3}$.
We devise a simplified CTRW model~\cite{HelfferichZiFrMeFaBlBa-PRE2014}, where we
assume that the tracer particles only move when there is an event close by.
In this picture the mean-square displacement is given by the number $N$ of events in a time
interval $t$ times the typical displacement during jumps.
The Laplace transform of $N(t)$ is related to the waiting time distribution as
$\tilde{N}(s)=\tilde{\psi}(s)/(s(1-\tilde{\psi}(s))$ and
is shown in the inset of Fig.~\ref{fig:statistical-properties}c.
We observe two regimes: one proportional to $s^{-2}$, this is, the long time diffusion,
and a second one proportional to $s^{-5/3}$, subdiffusive.
This leads to a prediction of $\langle\Delta r^2\rangle(t)\propto t^{1/3}$
that does not compare well with an exponent of about $0.85$ estimated from a power-law
fit in the subdiffusive regime corresponding to $T_1$ in Fig.\ref{fig:diffusion-and-structure-factor}(a).
We expect this to be due to the rough assumption of dynamical arrest between large jumps.
Still, the qualitative picture and the duration of the sub-diffusive regime are captured.
We deduce that this regime results from the negative correlations in the
displacements combined with the intermittent dynamics for the activity,
a feature that we like to call ``statistical caging''.

It is the local activity intermittency (power-law distributed) that allows
the emergence of a correlated dynamical regime, like the subdiffusive one observed here,
and gives rise to crossovers in time scale that will impact any measurement covering them.
For instance, coming back to the analysis of $S(q,t)$, we notice that during the sub-diffusive regime,
the relaxation time $\tau_r$
scales as $\tau_r \sim q^{-n}$ with $n>2$.
When rescaling time and wavelengths as $q^n t$ with the appropriate $n$, curves corresponding to a time window
where $D(t)$ decreases, collapse onto a new master curve $S(q,t)=\exp \left[-\alpha_s(q^nt)^\gamma \right]$, now with $\gamma<1$.
This anomalous diffusion with a stretched behavior of $S(q,t)$ is not accessible in a mean-field
approximation, and we could expect it \textit{a priori} to be realized in experiments.

{\it Discussion --}
Despite the strong simplifications we made to derive our model description,
we expect the qualitative features to be relevant in real systems.
We tested our model reproducing mean-field predictions for the distribution of the absolute
values of the tracer displacements and related values of the dimension-dependent
shape parameters in the decay of the dynamical structure factor $S(q,t)$.
We perfectly fit $S(q,t)$ in the ballistic regime
with a compressed exponential of shape parameter $\gamma \approx 1.8$ (given $\tau_f \propto q^{-1.125}$),
a value close to expected mean-field value in two dimensions $\gamma_\mathrm{MF}=2$.
We have observed (data not shown) that we further approach $\gamma \approx 2$ when we address smaller time
scales compared to the event duration by increasing $\tauOff$, and that this is accompanied by a clearer
ballistic ($\tau_f \propto q^{-1}$) scaling of the curves.
Note that the anomalous structural relaxation coexists with a stretched exponential decay of two
time autocorrelations in the local stresses [Fig.~\ref{fig:statistical-properties}(d)].

We insist that the commonly referenced $\gamma_\mathrm{MF}=3/2$ is valid in three dimensions only,
and for times smaller than the typical rearrangement duration.
We observe that the comparison between experiments and the mean-field prediction
is frequently inaccurate in the literature, failing in basic aspects as dimensionality
mismatching and/or overlooking the range of validity of the predictions.
Interestingly, the observed value $\gamma \approx 1.8$ coincides with experimental results on
effectively two-dimensional systems in the high-density and small $q$ regime~\cite{OrsiRuChPuRuBaRiCr-PRE2014}.
In that work, a $q$ dependence of the shape parameter is also reported.
We think that this feature is not captured by our model,
since it considers only point-like rearrangements and does not resolve the scales comparable to their size.
Experimental estimations of rearrangement typical size and duration
are fundamental to interpret the $q$-dependence of the measured shape parameter
and compare with theoretical predictions.
Such information is also indispensable to distinguish between a ballistic motion ruled by typical
displacements induced by a single rearrangement and one (yet not acknowledged in simulations)
caused by correlations among events instead.

We could claim at this point that even when aging is typically present in all experimental studies
reporting compressed exponentials, and actually affects the typical relaxation time, it is not
necessarily a key ingredient to observe this kind of phenomenology.
In fact, the same kind of relaxation has been reported very recently in a stationary
state~\cite{TamboriniCiRa-PRL2014}, free of aging.

Beyond the mean field results, we find that, at least in two-dimensional systems, correlations between events lead to
a partial confinement of the tracers generated by an evolving displacement field with in time anticorrelations.
This phenomenon is \textit{a priori} different from the traditional atomic caging effect, 
that happens at smaller length and time scales.
We call it ``statistical caging''.
Instead of enhancing the persistence of the tracer particles as often assumed in the literature,
correlations lead in our model to subdiffusive behavior.

We acknowledge financial support from ERC Grant No. ADG20110209.
J.-L.B. is supported by IUF.
Simulations on the CURIE hybrid cluster at TGCC were possible thanks to the GENCI Project No. t2014097236.
We thank L.~Cipelletti, B.~Ruta, E.~Bertin, A.~Rosso, D.~Vandembroucq and A.~Nicolas for useful discussions.

\bibliography{mesoplastic}
\bibliographystyle{apsrev4-1}
\end{document}